\documentstyle[prl,aps,multicol,epsfig,epsf,psfrag,amsmath]{revtex}

\large 
\begin{document}
\draft 
\title{
Current and noise in a model of an AC-STM molecule-metal junction
} 
\author{R. Guyon$^{1,2}$, T. Jonckheere$^{1}$, V. Mujica$^{3}$, A. Cr\'epieux $^{1,2}$  and
T. Martin$^{1,2}$ }
\address{ $^1$ Centre de Physique Th\'eorique,  
Case 907 Luminy, 13288 Marseille Cedex 9, France} 
\address{ $^2$ Universit\'e de la M\'editerran\'ee,  
13288 Marseille Cedex 9, France} 
\address{$^3$ Escuela de Qu\'{\i}mica, Facultad de Ciencias, Universidad Central de Venezuela, Apartado 47102, Caracas 1020A, Venezuela}

\maketitle 
\begin{abstract} 

The transport properties of a simple model for a finite level structure (a {\it molecule} or a dot) connected to metal electrodes in an 
alternating current scanning tunneling microscope (AC-STM) configuration is studied.
The finite level structure is assumed to have strong binding properties with the metallic 
substrate, and the bias between the STM tip  and the hybrid metal-{\it molecule} interface has both an AC 
and a DC component. The finite frequency current response and the zero frequency 
photo-assisted shot noise are computed using the Keldysh technique, and examples for a single site {\it molecule} (a quantum dot) and for a two-site {\it molecule} are examined. The model may be useful for the interpretation of recent experiments using an AC-STM for the study of both conducting and insulating surfaces, where the third harmonic component of the current is measured.  The zero frequency photo-assisted shot noise serves as a  useful diagnosis for analyzing the energy level structure of the {\it molecule}.  The present work motivates the need for 
further analysis of current fluctuations in electronic molecular transport.   
     

 \end{abstract}

\begin{multicols}{2}

\section{Introduction}

The understanding of electronic transport phenomena in molecule-electrode 
interfaces is a key element of any technological
realization of the potential of single-molecule electronics\cite{ratner,joachim}. 
The basic physics governing the current voltage characteristics 
in the coherent regime is reasonably well understood in terms of generalizations of the Landauer model 
where current is viewed as a quantum scattering problem and the effect of the electrodes is included 
via a self-energy contribution. Such a model has been successfully used to interpret data from both
isolated molecules in STM, AFM and break junctions experiments\cite{reichert,bumm,kergueris}, as well as 
experiments on Self Assembled Monolayers\cite{datta1,ulman,dumas}. Most experiments on molecular-scale devices have been carried
out on a two-terminal assembly, but measurements have also been reported in gated systems\cite{zhitenev,park}. 
A vast majority of experimental and theoretical studies have focused on the measurement of the current 
in such molecular junctions. Generally speaking, in the coherent tunneling regime,
the zero frequency current $I(0)$ in molecular wire 
junctions depends on the 
electronic structure of the extended {\it molecule}-electrode system, the spatial profile of the voltage
drop, the electro-chemical potential across the junction and the self-energy of the contacts that
contains the information about the coupling between the finite {\it molecule} and the extended electrode\cite{emberly,xue}. 
Yet from the point of view of mesoscopic physics, the current alone is not sufficient 
to fully characterize the transport: the second moment of the current -- the noise --
has been shown to provide crucial information concerning the effective charge of the carriers
and their statistics\cite{blanter,chen}. While noise measurements still represent a certain challenge
in molecular transport -- as experiments are typically performed with tunnel contacts --
recently noise was computed for simple molecules using a H\"uckel type Hamiltonian 
\cite{walczak,dallakyan} and in the case of Coulomb interaction in the molecule 
using an Anderson impurity model\cite{thielmann}. 
The present work deals with the computation of both current and noise when
the bias imposed across the {\it molecule} has both a constant (DC) and an alternating (AC) component\cite{bruder,kouwenhoven}. 

Indeed, the alternating current STM (AC-STM) was suggested by Kochanski \cite{kochanski} as a 
experimental technique which could extend the capabilities of conventional STM from conducting to 
insulating surfaces.  A recent experiment\cite{blum} on the oxidation of lead sulfide surfaces
confirmed that crucial information could be obtained from the analysis of the third harmonic 
of the current. Nevertheless, there has been so far no report of an AC--STM diagnosis on a Self--Assembled Monolayers as a means to probe single molecule transport.  

From the point of view of the theoretical community, the problem of electron transport in the presence of 
both a DC and an AC bias voltage is often referred to as photo-assisted transport \cite{aguado,schoelkopf2,levinson,pedersen}.
In the absence of an AC bias, most treatments of molecular wire junctions focus 
on the problem of stationary current and conductance. On the other hand, the problem of AC-STM requires an explicit
time-dependent approach that allows a full description of the transient terms, as well as the stationary terms. 
In this regard a Hamiltonian formulation of transport such as the Keldysh method\cite{mii,jauho,heurich} 
is more powerful than the scattering-based Landauer approach\cite{mujica}. 
It allows a proper treatment of the electron statistics in real time.

Here a simple model junction is considered, which consists of two metal electrodes and a 
finite electronic structure, either a dot or a {\it molecule} described
by a tight-binding Hamiltonian, which is coupled to the electrodes through 
a single contact. The system is subject to the combined action
of a DC voltage bias and a time-varying field. 
The Keldysh formulation allows to compute both the 
frequency-dependent current response as well as the photo--assisted shot noise. The {\it molecule} is 
assumed to be well connected to one of the electrodes (the substrate) and its binding properties are 
computed non perturbatively. The coupling to the other electrode (STM tip) is treated to second order 
in the hopping amplitude, but the field amplitude appears to all orders in this problem as will be shown below.  
Under simplifying assumptions (Poisson regime), the noise is just proportional to the average 
stationary current itself, and therefore both contain essentially the same information. 
This Shottky relation between current and noise ceases to be 
valid for higher order harmonics of the current and therefore the study of the noise can 
reveal important informations about a periodically driven system. 
Finally, one should emphasize that the general method employed here is in no sense 
specific to a metal/single-dot/metal junction. Here one has chosen to proceed with 
specific computations on this particular system, but additional results on the two site {\it molecule}
can be generalized to any linear molecular chain. 

This article is organized as follows: Section 2 presents a description of the model. Section 3 is devoted to the computation of non--equilibrium transport quantities. In section 4, the main results for the photo--assisted shot noise for both the case of a single dot {\it molecule} and the two--site {\it molecule} are shown and discussed. The higher harmonics of the current are examined in section 5. Finally, section 6 contains some concluding remarks and speculations.

\section{Model Hamiltonian}
\vskip1cm

\begin{figure}  
\epsfxsize 4 cm  
\centerline{\epsffile{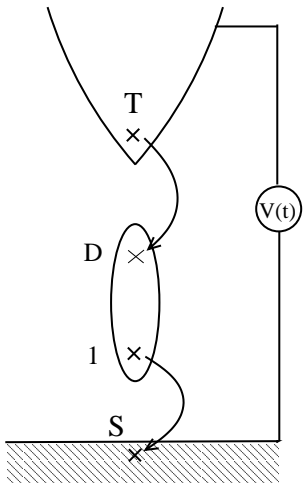}}  
\medskip  
\caption{\label{rodolphe1}
System : the {\it molecule} (quantum dot) sandwiched between a substrate and an STM tip.}   
\end{figure} 

Consider the model for a  metal--{\it molecule}--metal junction depicted in Fig. \ref{rodolphe1}.
The substrate and the tip are assumed to be metallic and are both described by an infinite 
bandwidth model. Both are therefore characterized by constant density of states 
$\rho_S$ and $\rho_T$.

In turn, the {\it molecule} will consist either of a single dot or of a conjugated structure
(described by a tight binding Hamiltonian). It contains sites numbered from $1$ to $D$. The 
connection between the {\it molecule} and the substrate occurs on site $1$, while
the connection between the {\it molecule} and the tip is located at site $D$.
This general approach allows to compute the transport properties for 
a {\it molecule} with an arbitrary number of sites, but specific plots will be obtained for 
$D=1$ and $D=2$. 

To be more specific,  
the connection between the {\it molecule} and the substrate is assumed to be quite strong.
Indeed, such strong binding to a substrate is known to occur when the {\it molecule} is 
functionalized: when for instance an end group of atoms in the {\it molecule} is replaced 
by a sulphur atom, which has strong binding properties on a gold substrate.

When the other electrode (the STM tip) is brought close to the {\it molecule}, and a bias 
voltage is imposed between both metallic electrodes,
non equilibrium flow of electrons occurs. 
The key point of this work is to use an alternating current STM for probing 
transport. This is modeled by adding 
an alternating contribution to the DC bias $V_0$ imposed on the junction: 
$V(t) = V_0 + V_1 cos(\omega t)$ where $V_{1}$ is the amplitude of the
AC bias voltage and $\omega$ is its frequency. 

The sites involved in the junction 
between the {\it molecule} and the tip are denoted $D$ (as in dot) and $T$ (tip).
With this convention, the Hamiltonian  describing tunneling through this 
junction reads:
\begin{equation}
H_{T D}(t) = \Gamma(t) c^\dag_T   c_D + \Gamma^*(t) c^\dag_D   c_T 
\end{equation}
where $c/c^\dag$ are fermionic annihilation/creation operators.
The time dependence of the  voltage bias is thus included in the hopping term using the  
Peierls substitution: 
starting from a formulation where the electric field at the junction is described
solely by a scalar potential, a gauge transformation with gauge parameter 
$e \chi(t) c= - e \int V(t) dt$ is performed.
The phase of the amplitude has a linear dependence associated with the 
DC bias, but it also contains a modulation due to the AC voltage:
\begin{eqnarray}
\Gamma(t) &=& \Gamma^0_T \exp[- i e \chi(t) / c]\nonumber\\
&=& \Gamma^0_T \exp(i \left(\omega_0 t + \frac{\omega_1}{\omega} sin(\omega t) \right))
\end{eqnarray}
where $\Gamma^0_T$ is the time independent tunneling amplitude
of the hopping between the tip and the {\it molecule}, and $\omega_{0,1}\equiv eV_{0,1}$.

An advantage of the Peierls substitution is that the current operator can be directly obtained by 
differentiating the hopping Hamiltonian with respect to the gauge parameter:
\begin{equation} 
\hat{I}(t) = -c  \frac{\partial H_{TD}(t)}{\partial \chi(t)}
\label{current_definition}
\end{equation}

Moreover, the current--current correlations (the current noise) are defined by 
\begin{equation}
S(t,t') ={1 \over 2} \left[\langle \hat{I} (t) \hat{I}(t')
+ \hat{I}(t') \hat{I}(t) \rangle - \langle \hat{I}(t) \rangle 
\langle \hat{I} (t') \rangle\right]\label{noise_definition}
\end{equation}
Note that this correlator is symmetrized, which is appropriate when computing zero frequency noise: when considering the finite frequency spectral
density of noise, this ceases to be true and one needs to consider the unsymmetrized correlator, 
which makes the difference between emission or absorption \cite{aguado}.

\section{Non-equilibrium transport}

Note that because of the time dependent bias, any straightforward extension of the 
Landauer approach has to be ruled out, and the Keldysh formalism
is employed. 

The expectation value of the current (Eq.~(\ref{current_definition})) can be directly cast in terms of the spectral 
Green's functions. For the noise (Eq.~(\ref{noise_definition})), we can factorize the two body correlation function following a mean 
field decoupling scheme, which is exact for non-interacting electrons.
Green's functions are represented by $2\times 2$ matrices:
each time argument can be pinned on either branch of the Keldysh time contour.
However, recall that these Green's functions are not independent.  
For convenience, here, one chooses to use the version of Dyson equations for advanced, 
retarded and spectral Green's functions : 
\begin{eqnarray}
G^a_{\alpha \beta}(t,t') &=& i \, \Theta(t'-t) \langle \{  c_{\alpha} (t) , c_{\beta}^+ (t')\} \rangle \\
G^r_{\alpha \beta}(t,t') &=& -i \, \Theta(t-t') \langle \{  c_{\alpha} (t),  c_{\beta}^+ (t') \} \rangle \\
G^{+-}_{\alpha \beta}(t,t') &=& i \langle c^\dag_{\beta} (t') c_{\alpha}(t) \rangle \\ 
G^{-+}_{\alpha \beta}(t,t') &=& -i \langle c_{\alpha} (t) c^\dag_{\beta}(t') \rangle
\end{eqnarray}
Here $\alpha, \beta$ are terminal indices.
Inserting these Green's functions in the current and the real time current-current correlator yields:
\begin{eqnarray}
I(t) &=& 2 e  Re\left[ \Gamma(t)  G^{+-}_{D T}(t,t) \right]
\nonumber \\
S(t,t') &=&  2 e^2   Re\Big[  \Gamma(t)  \Gamma^* (t') G^{+-}_{TT}(t',t) G^{-+}_{DD}(t,t')) \nonumber \\
&&~~~~~~~~ +  \Gamma^*(t)  \Gamma (t') G^{+-}_{DD}(t',t) G^{-+}_{TT}(t,t') \Big]
 \end{eqnarray}
Here one is only interested in the lowest order contribution in the tip--{\it molecule}
hopping amplitude.  The Dyson equation for the above Green's function reads
(omitting terminal indices summations and time integrations):
\begin{eqnarray}
G^{a,r} &=& g^{a,r} (1 + \Sigma^{a,r} G^{a,r}) \nonumber \\
G^{+-,-+} &=& (1+G^r  \Sigma^r) g^{+-,-+} (1+\Sigma^a G^a  )
\end{eqnarray}
In the perturbative scheme, the  $G$'s are the dressed Green's functions -which have both diagonal and non-diagonal non-zero
elements in the terminal indices- while the $g$'s are the bare Green's functions with only diagonal non-vanishing elements. 
The self-energies $\Sigma$ are related to the hopping amplitude : $\Sigma^{r} = \Gamma(t)$,  $\Sigma^{a} = \Gamma^{*}(t)$,
$\Sigma^{+-,-+} = 0$. 
At this point the time-dependent hopping amplitude is expanded using the generating function 
of the Bessel functions\cite{land_butti} $\Gamma(t) = \Gamma^0_T \sum_{n=-\infty}^{+\infty} \; e^{i(\omega_0 + n \omega)t} 
J_n\left(\frac{\omega_1}{\omega} \right)$.

The strong coupling between the substrate and the {\it molecule} allows to define a 
compound {\it molecule}--substrate system. The Green's function of the {\it molecule} becomes dressed
by the substrate-{\it molecule} coupling, which can be treated non perturbatively.
In the following, $\tilde{G}$ denotes the non-equilibrium dressed Green's 
functions for this composite system. 

A Fourier transform ($h(\omega) = \int_{-\infty}^{+\infty} dt e^{i \omega t} h(t)$) helps to
characterize the current response at finite frequency. Because time-translational invariance 
is broken by the AC bias, a double Fourier transform is implied for the noise:
\begin{eqnarray}
I(\Omega) &=& e |\Gamma^0_T|^2 \sum_{n,m} 
J_n \left({\omega_1 \over \omega} \right)  
J_m \left({\omega_1 \over \omega} \right) \int_{-\infty}^{+\infty} d\omega_2
\nonumber \\
\times \Big\{ &&  
 [\delta(\Omega +(n-m)\omega) + \delta(\Omega -(n-m)\omega)] \nonumber \\
&&~~~~~~~~~~~~\times \tilde{G}^{+-}_{DD}(\omega_2) g^a_{TT}(\omega_2 - \omega_0 -m \omega) \nonumber \\
&+& \delta(\Omega +(n-m)\omega)\nonumber \\
&& \times \Big[ \tilde{G}^{r}_{DD}(\omega_2) g^{+-}_{TT}(\omega_2 - \omega_0 -m \omega)\nonumber \\
&&~~~~~~ -  \tilde{G}^{a}_{DD}(\omega_2) g^{+-}_{TT
}(\omega_2 - \omega_0 -n \omega)
\Big] \Big\}
\nonumber \\
S(\Omega,\Omega') &=& 
 e^2 |\Gamma^0_T|^2 \sum_{n,m} J_n \left({\omega_1 \over \omega} \right)  J_m \left({\omega_1 \over \omega} \right)
\nonumber \\
&&~~ \times \delta(\Omega + \Omega' +(n-m)\omega)  \int_{-\infty}^{+\infty} d\omega_2
\nonumber \\
\times \Big\{   \tilde{G}^{-+}_{DD}(\omega_2) &&
 \Big[  g^{+-}_{TT}(\omega_2 +\Omega' - \omega_0 -m \omega) \nonumber \\
&&~~~~~~~+  g^{+-}_{TT}(\omega_2 - \Omega' - \omega_0 -n \omega) \Big]  \nonumber \\
+ \tilde{G}^{+-}_{DD} (\omega_2) && \Big[  g^{-+}_{TT}(\omega_2 +\Omega - \omega_0 -m \omega)  \nonumber \\
&&~~~~~~~
+  g^{-+}_{TT}(\omega_2 -\Omega - \omega_0 -n \omega) \Big]  \Big\}
\end{eqnarray}

At this point, it is convenient to introduce the tip's Green's functions (Appendix \ref{appelectrode}) where $\rho_T$ is the 
density of states in the tip. The noise and the current thus become:
\begin{eqnarray}
I(\Omega) &=& i e \pi \rho_T  |\Gamma^0_T|^2 \sum_{n,m} J_n \left({\omega_1 \over \omega} \right)  J_m \left({\omega_1 \over \omega} \right)\nonumber \\
\times \Big\{ && 
[\delta(\Omega +(n-m)\omega) + \delta(\Omega -(n-m)\omega)] \nonumber \\
&& \times 
\int_{-\infty}^{+\infty} d\omega_2 \tilde{G}^{+-}_{DD}(\omega_2) \nonumber \\
+ 2 && \delta(\Omega +(n-m)\omega) \nonumber \\
&& \times \Big[ \int_{-\infty}^{\omega_0+m \omega}
 d\omega_2 \left(\tilde{G}^{r}_{DD}(\omega_2) -  \tilde{G}^{a}_{DD}(\omega_2) \right)\nonumber \\
&&~~~~~~ -  \int_{\omega_0 + m \omega}^{\omega_0 + n \omega}  d\omega_2  \tilde{G}^{a}_{DD}(\omega_2) \Big] \Big\}
\label{current_dressed}\end{eqnarray}

\begin{eqnarray}
S(\Omega,\Omega') &=& 
 (2 e) i e \pi \rho_T   |\Gamma^0_T|^2 \sum_{n,m} J_n \left({\omega_1 \over \omega} \right)  J_m \left({\omega_1 \over \omega} \right) 
\nonumber \\
&& \times \delta(\Omega + \Omega' +(n-m)\omega) 
\nonumber \\
\times \Big\{ && -2 \int_{\omega_0 + m\omega - \Omega}^{+\infty} d\omega_2 \;  \tilde{G}^{+-}_{DD}(\omega_2) \nonumber \\
&& + 2 \int_{-\infty}^{\omega_0 + m\omega - \Omega} d\omega_2 \tilde{G}^{-+}_{DD}(\omega_2)\nonumber \\
&& +  \int_{\omega_0 + m\omega - \Omega}^{\omega_0 + n\omega + \Omega} d\omega_2
 \left[  \tilde{G}^{-+}_{DD}(\omega_2) + \tilde{G}^{+-}_{DD}(\omega_2) \right] \Big\}\nonumber \\
\label{noise_dressed}
\end{eqnarray}
The noise and current are fully determined once the 
dressed Green's functions of the {\it molecule} (dot) are specified. 
Note that Eqs. (\ref{current_dressed}) and (\ref{noise_dressed}) are 
valid for an arbitrary model of the {\it molecule}/substrate, which could in 
principle include interactions. The $\delta$ functions in Eq. (\ref{current_dressed}) and (\ref{noise_dressed}) restrict the frequency dependance of the current to integer multiple of the driving frequency $\omega$. The {\it m}th harmonic of the current, $I(m \omega)$, is obtained by integrating Eq. (\ref{current_dressed}) on the detector resolution $s$ : $I(m \omega) = \int_{m \omega - s /2}^{m \omega + s /2} d\Omega I(\Omega)$, then getting rid of the $\delta$ function singularities. The detector resolution $s$ is of course much smaller than all the frequencies of this problem.

Here we focus on a single dot and on a two site {\it molecule}.
The dressed Green's functions are derived in Appendices \ref{appdressed} and \ref{appdressed1}.
For the single dot they read:
\begin{eqnarray}
\tilde{G}^{a,r}_{DD}(\Omega) &=&  \frac{1}{ \Omega -E_D \mp i   \pi \rho_S  \Gamma_S^2}\nonumber \\
\tilde{G}^{+-}_{DD}(\Omega) &=& 
2 i \pi \rho_S  \Gamma_S^2
\frac{ f(\Omega-E_F)
}{
(\Omega - E_D)^2 +( \pi \rho_S  \Gamma_S^2
)^2} \nonumber \\
\tilde{G}^{-+}_{DD}(\Omega) &=& - 2 i  \pi \rho_S  \Gamma_S^2
\frac{ (1 - f(\Omega-E_F))
}{
(\Omega - E_D)^2 +( \pi \rho_S  \Gamma_S^2
)^2} \nonumber \\
\end{eqnarray}

The coupling to the metallic substrate imposes a width
$\pi \Gamma_S^2\rho_S$. Furthermore, the occupation of the dot 
(as it appears in  $\tilde{G}^{+-}_{DD}$) is specified by the Fermi function $f$
of the substrate. 

For two site {\it molecule}, the dressed 
Green's functions are derived in Appendix \ref{appdressed1} :

\begin{eqnarray}
\tilde{G}^{a,r}_{DD}(\Omega) &=& -  \frac{ \Omega - i\pi \rho_S \Gamma_S^2} {t^2 - \Omega (\Omega -  i \pi \rho_S \Gamma_S^2)} \nonumber \\
\tilde{G}^{+-}_{DD}(\Omega) &=& 2 i \pi \rho_S \Gamma_S^2\frac{t^2 f(\Omega - E_F)}{|t^2 - \Omega (\Omega +  i \pi \rho_S \Gamma_S^2)|^2} \nonumber \\
\tilde{G}^{-+}_{DD}(\Omega) &=&- 2 i \pi \rho_S \Gamma_S^2\frac{t^2 [ 1- f(\Omega - E_F)]}{|t^2 - \Omega (\Omega +  i \pi \rho_S \Gamma_S^2)|^2}
\end{eqnarray}

where $t$ is the hopping between the two sites of the {\it molecule}. 

Note that, in both cases (single dot/ double dot), the spectral dressed Green functions may be written in terms of the density of states of the dressed system :

\begin{eqnarray}
\label{density}
\tilde{G}^{+-}_{DD}(\Omega) &=& 2 i Im[\tilde{G}^{a}_{DD}(\Omega)] f(\Omega - E_F) \nonumber \\
 \tilde{G}^{-+}_{DD}(\Omega) &=& -2 i Im[\tilde{G}^{a}_{DD}(\Omega)] [1-f(\Omega - E_F)] \nonumber \\
 \end{eqnarray}

where $Im[\tilde{G}^{a}_{DD}(\Omega)]/ \pi$ is  the density of states on site $D$ .  This density should be computed directly from the expression of the advanced dressed Green's function :

\begin{eqnarray}
\label{density1}
Im[\tilde{G}^{a}_{DD}(\Omega)]|_{1 dot} &=& \frac{ \pi \rho_S \Gamma_S^2}{(\Omega - E_D)^2 + (\pi \rho_S \Gamma_S^2)^2} \nonumber \\
Im[\tilde{G}^{a}_{DD}(\Omega)]|_{2 dots} &=& \frac{ \pi \rho_S \Gamma_S^2 t^2}{(t^2 - \Omega^2)^2 + (\pi \rho_S \Gamma_S^2)^2 \Omega^2}
\end{eqnarray}

In the following, the integrated version of the density of states denoted by $F$ is needed :

\begin{equation}
\label{F}
F(a,b) = \int_{a}^{b}d u \;Im[\tilde{G}^a_{DD}(u)]~.
\end{equation}

\section{Photo-assisted shot noise}

One is now in a position to compute the current harmonics and the 
photo-assisted shot noise using the explicit expressions for the 
dressed Green's functions. Before displaying these results, 
as a reference it is 
useful to recall what happens in a tunnel junction between two 
metallic electrodes (no {\it molecule}). This allows to relate to the 
results of Ref. \onlinecite{lesovik_levitov}, where 
photo-assisted shot noise was studied in a mesoscopic conductor from the 
point of view of scattering theory. For the current and the noise our perturbation scheme yields at zero frequencies :
\begin{eqnarray}
I(\Omega = 0) &=& 4 e \pi^2 \rho_T \rho_S   \Gamma_S^2  \sum_n 
J_n^2 \left( {\omega_1 \over \omega } \right) (\omega_0 + n\omega -E_F)\nonumber \\
S(0,0) &=& (2 e) 4 e \pi^2 \rho_T  \rho_S    \Gamma_S^2  \sum_n 
J_n^2 \left( {\omega_1 \over \omega } \right) |\omega_0 + n\omega -E_F|\nonumber \\
\end{eqnarray}
The noise is continuous, but its derivative has jumps at integer values of 
$\omega_0/\omega$, precisely the result of Ref. \onlinecite{lesovik_levitov}
when one assumes that the transmission coefficient used for scattering theory 
does not depend on energy. It is therefore relevant to generalize 
these results to a situation where the ``mesoscopic system'' has 
an internal  discrete level structure, which is the case for molecular 
electronic transport. This is expected to have 
repercussions on both the current harmonics and the noise.     

Using the general formulation of the spectral Green's functions (Eq.~\ref{density}), the zero frequencies noise read :

\begin{eqnarray}
\label{noise_gen}
 S(0,0) &=&  8 e^2 \pi \rho_T  |\Gamma^0_T|^2 \sum_n \bigg[
J_n^2 \left( {\omega_1 \over \omega } \right) sgn(\omega_0 + n \omega - E_F)\nonumber \\
&&~\times F(E_F,\omega_0 + n \omega)\bigg]
\end{eqnarray}

where $F(a,b)$ is defined in Eq. (\ref{F}).

In order to make a diagnosis on the spectrum, it is useful to display 
also the first derivative of the noise with respect to the DC voltage:

\begin{eqnarray}
\label{noise_der_gen}
\frac{dS(0,0)}{d\omega_0} &=&  8 e^2 \pi \rho_T  |\Gamma^0_T|^2  \sum_n \bigg[ 
J_n^2 \left( {\omega_1 \over \omega } \right) sgn(\omega_0 + n \omega - E_F)\nonumber \\
&&~\times  Im[\tilde{G}^a_{DD}(\omega_0 + n \omega)] \bigg]\nonumber \\
\end{eqnarray}

\subsection{Single dot problem}

One of the motivations for studying photo-assisted shot noise is 
to probe the finite frequency spectral density of noise using the external 
frequency $\omega$ of the AC bias. We will therefore focus on the photo-assisted shot noise 
with both frequencies in Eq. (\ref{noise_dressed}) set to zero, because 
low frequency measurements (still above, say  $100KHz$, to avoid $1/f$ noise) are more 
accessible experimentally than high frequency ones.  Using the definition of the density of states for a single dot system (Eq. (\ref{density1})) and after performing the integration over frequencies in Eqs. (\ref{noise_gen}) and  (\ref{noise_der_gen}), the noise at zero frequencies and its derivative read:

\begin{eqnarray}
&& S(0,0) =  8 e^2 \pi \rho_T  |\Gamma^0_T|^2 \sum_n 
J_n^2 \left( {\omega_1 \over \omega } \right) sgn(\omega_0 + n \omega - E_F)\nonumber \\
&&\times \left[ \arctan\left(
{(E_D - E_F) \over \pi \rho_S  \Gamma_S^2}
\right)
- \arctan\left(
{(E_D - \omega_0 - n \omega ) \over \pi \rho_S  \Gamma_S^2 }
\right) \right]\nonumber \\
\label{noise_1dot}\end{eqnarray}
\begin{eqnarray}
\label{noise_derivative_1dot}
 \frac{dS(0,0)}{d\omega_0} &=& 8 e^2 \pi^2 \rho_T\rho_S   \Gamma_S^2 |\Gamma^0_T|^2   \sum_n 
J_n^2 \left( {\omega_1 \over \omega } \right) \nonumber \\
&&~\times  \frac{sgn(\omega_0 + n \omega - E_F)}{
(\omega_0 + n \omega -E_D)^2 + (\pi \rho_S  \Gamma_S^2)^2
}
\end{eqnarray}
\vskip1cm

\begin{figure}
\psfrag{s00}{$S(0,0)$}
\psfrag{ds00tototttttttttt}{${dS(0,0)}/{d \omega_0}$}
\psfrag{omega0}{${\omega_0}/{\omega}$}
\psfrag{2electrodesgggggg}{2 electrodes}
\psfrag{gs0.1}{$\Gamma_S = 0.1$}
\psfrag{gs1.0}{$\Gamma_S = 1.0$}
\psfrag{gs10}{$\Gamma_S = 10.$}
\psfrag{ef}{$E_F$}
\centerline{\includegraphics[width=8.cm]{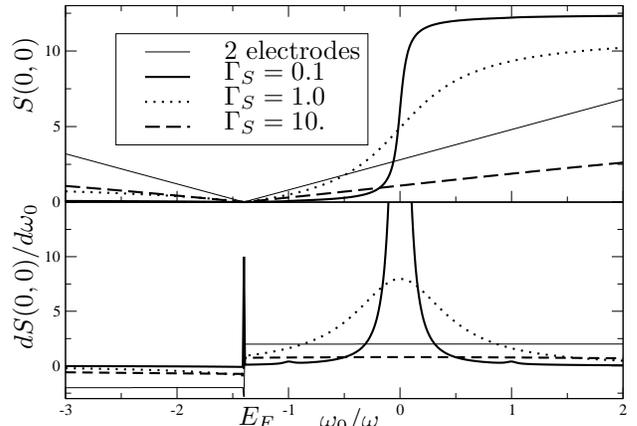}}
\caption{\label{petitac}Zero--frequencies noise $S(0,0)$ (upper graph) 
and its first derivative $d S(0,0) / d \omega_0$ (lower graph) as a function 
of the ratio $\omega_0/\omega$ in the regime of a small AC bias ($\omega_1 / \omega = 0.1$) for the single dot system. Fermi energy of the substrate $E_F$ is 
fixed to $-1.4 \omega $ and energy level of the dot $E_D$ is the origin of the energy. 
Comparisons are made between the two electrodes case and different values of 
hopping $\Gamma_S$ in the dot--metal junction. }
\end{figure}

\begin{figure}
\psfrag{s00}{$S(0,0)$}
\psfrag{ds00tototttttttttt}{${dS(0,0)}/{d \omega_0}$}
\psfrag{omega0}{${\omega_0}/{\omega}$}
\psfrag{2electrodes}{2 electrodes}
\psfrag{gs0.1}{$\Gamma_S = 0.1$}
\psfrag{gs1.0}{$\Gamma_S = 1.0$}
\psfrag{gs10}{$\Gamma_S = 10.$}
\psfrag{ef}{$E_F$}
\centerline{\includegraphics[width=8.cm]{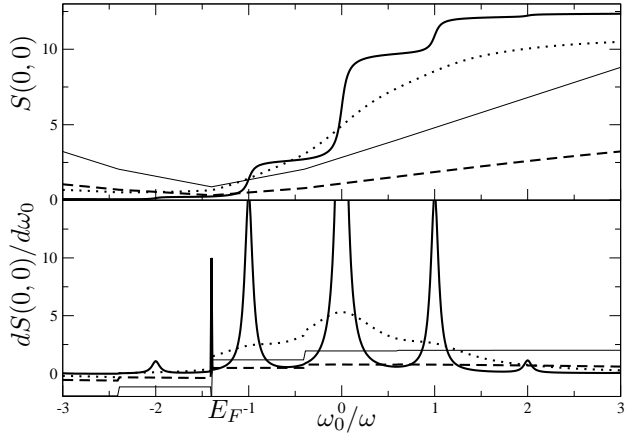}}
\caption{\label{moyenac} Same as Fig.\ref{petitac} for an AC bias: $\omega_1 / \omega = 1.0$.}
\end{figure}

\begin{figure}
\psfrag{s00}{$S(0,0)$}
\psfrag{ds00tototttttttttt}{${dS(0,0)}/{d \omega_0}$}
\psfrag{omega0}{${\omega_0}/{\omega}$}
\psfrag{2electrodes}{2 electrodes}
\psfrag{gs0.1}{$\Gamma_S = 0.1$}
\psfrag{gs1.0}{$\Gamma_S = 1.0$}
\psfrag{gs10}{$\Gamma_S = 10.$}
\psfrag{ef}{$E_F$}
\centerline{\includegraphics[width=8.cm]{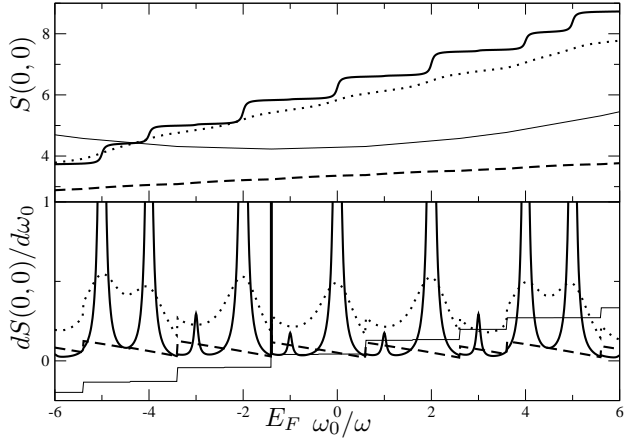}}
\caption{\label{grandac} Same as Fig.\ref{petitac} for an AC bias: $\omega_1 / \omega = 10.0$.}
\end{figure}

\begin{figure}
\psfrag{s00}{$S(0,0)$}
\psfrag{ds00}{${dS(0,0)}/{d \omega_0}$}
\psfrag{ds00tototttttttttt}{${dS(0,0)}/{d \omega_0}$}
\psfrag{omega0}{${\omega_0}/{\omega}$}
\psfrag{2electrodes}{2 electrodes}
\psfrag{gs0.1}{$\Gamma_S = 0.1$}
\psfrag{gs1.0}{$\Gamma_S = 1.0$}
\psfrag{gs10}{$\Gamma_S = 10.$}
\psfrag{ef}{$E_F$}
\centerline{\includegraphics[width=8.cm]{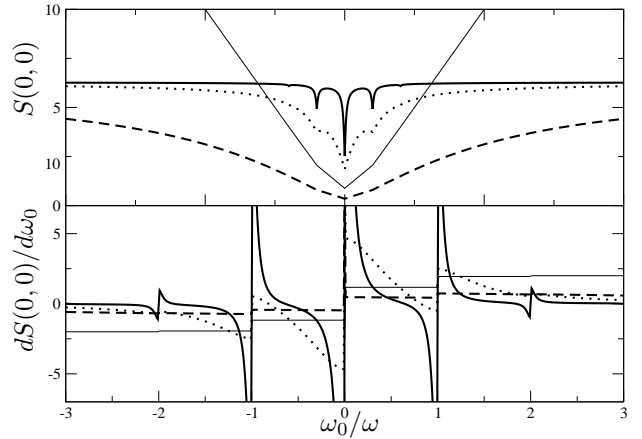}}
\caption{\label{moyenacres} Same as Fig.\ref{moyenac} in the case of a resonance 
between the energy level of the dot and the Fermi energy of the substrate: $E_F = E_D = 0$.}
\end{figure}

The quantities which are plotted on Figs \ref{petitac} to \ref{noise_2_dots} are normalized according to the prefactors which appear on Eqs. (\ref{noise_1dot}) and (\ref{noise_derivative_1dot}).

Consider first the case where the dot level lies above the Fermi level 
of the substrate.
Fig. \ref{petitac} displays the noise and its first derivative with respect to the DC bias
for a small value of the AC bias amplitude. 
 From the point of view of our analytical results,
this means that very few terms contribute to the sum over integers in Eq. (\ref{noise_1dot}):
$n=0$ dominates. In the case of large broadening, we recover similar results to the two 
electrodes system. The dot is metallized to the extent that transport is 
fully specified by the two Fermi distributions of the electrodes. 
As the broadening decreases, the noise acquires a step-like structure
close to the bare dot level, which gets sharper for lower  $\Gamma_S$. This is associated to 
the fact that the current increases drastically at this particular DC bias. Far from this 
DC bias value, the noise saturates because the AC bias does not allow for the exchange 
of ``photon'' quanta $ \omega$.  The peak appearing in the derivative of the noise for the case of a two electrodes system is a consequence of the abrupt change of sign of the current.

\vskip1cm  

When one further increases the AC bias (Fig. \ref{moyenac}), photonic exchanges become relevant: this is illustrated by the 
appearance of new shoulders in the noise (peaks in the noise derivative) for integer values of the ratio 
$\omega_0/\omega$. The dominant peak in the derivative is located at the dot level $\omega_0=0$. Side peaks
at $\omega_0=\pm \omega $ are clearly visible, while  those at  $\omega_0=\pm 2 \omega $ can only be identified 
with the help of the noise derivative.  These peaks are smoothed 
when the substrate coupling is increased, eventually leading to metallization as before.
\vskip1cm  

Turning now to the case of a large AC bias (Fig. \ref{grandac}), one notices that the step like structure is preserved but the
width of the steps is apparently doubled compared to Fig. \ref{moyenac}. This is only apparent, as the noise derivative 
displays secondary peaks associated with odd photon absorption/emission. 
 The relative importance of odd and  even photon peaks can be traced to a high sensitivity  
of the noise with respect to the AC bias amplitude $\omega_1$, due to the oscillations of the Bessel functions 
(see Eq.~(\ref{noise_derivative_1dot})).
\vskip1cm  

The behavior of this system turns out to be strongly dependent on the location of the dot level.
In Fig. \ref{moyenacres}, the dot level matches the Fermi level of the substrate.
The AC bias amplitude has been adjusted as in Fig. \ref{moyenac} where essentially 
$\omega_0=0,\pm\omega$ give salliant features.
First, notice that the noise does not display steps any more. 
Qualitatively, it is understood that current flows  
when the absolute value of the bias is large compared to the
the ``photon energy'' $ \omega$: in this situation there is no balance between carriers 
emmitted from the tip to the substrate, and vice versa. 
Next, the current -- and thus the noise -- must vanish when 
the tip and the substrate have matching Fermi energies.
More interesting is the reduction of the noise observed when $\omega_0=\pm\omega$.
Current can flow with the help of virtual transitions through the dot, but 
for this AC bias where essentially only single photon transitions contribute, there is an 
additional channel which brings about noise reduction. 
This explains the two side dips in the noise
at $\omega_0=\pm\omega$ in  Fig. \ref{moyenacres}. 
It is interesting to note that the behavior of the noise and of its derivative 
is drastically modified only when the coupling to the substrate is small, 
whereas for a metallized dot the typical behavior associated with two metallic 
electrodes is recovered.

\subsection{Two--site molecule}

The expressions for the dressed Green's 
functions of the two--site {\it molecule} are inserted in 
the expressions for the zero--frequencies noise. 
The two--site {\it molecule} has a hopping amplitude $t$ between
the two sites, so the bonding and anti-bonding orbitals 
are separated by $2t$.  Note that when the Fermi level
of both the tip and the substrate lie between these orbitals, 
these two level can mimic the HOMO (Highest Occupied Molecular Orbital)
and the LUMO (Lowest Unoccupied Molecular Orbital) of a more complex {\it molecule}.
Here one is interested in understanding how the presence of the two sites 
modifies the noise with respect to the single dot case.
Expressions for the noise and its derivative are given in Eqs.~(\ref{noise_gen}) and (\ref{noise_der_gen}) with the corresponding density of states (see Eq.~(\ref{density1})).

\vskip1.2cm

\begin{figure}
\psfrag{S00}{$S(0,0)$}
\psfrag{omega0}{${\omega_0}/{\omega}$}
\psfrag{a/}{a)}
\psfrag{b/}{b)}
\centerline{\includegraphics[width=8.cm]{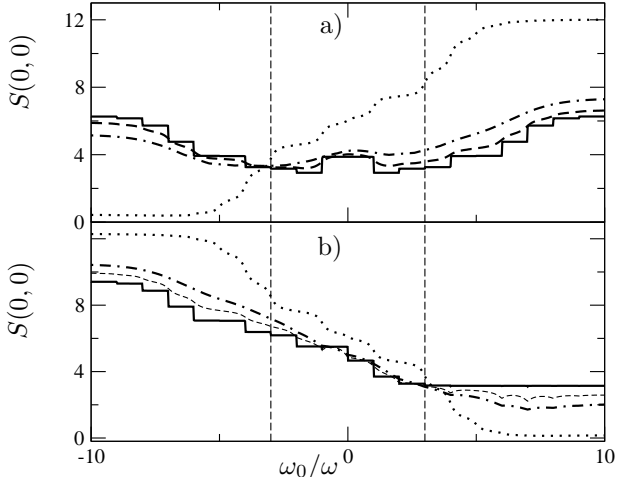}}
\caption{\label{noise_2_dots} Zero--frequencies noise $S(0,0)$ for a two--site {\it molecule} as a function 
of the ratio $\omega_0/\omega$ for AC bias $\omega_1 / \omega = 5.0$. Full line , $\Gamma_S = 0.001$; dashed line, $\Gamma_S = 1.0$;
dashed-dotted,  $\Gamma_S = 3.0$; dotted line, $\Gamma_S = 100$. The hopping between the 
two sites is $t = 3\omega$. Vertical dashed line correspond to the bonding and antibonding orbitals. 
Plots are given for:
a) $E_F = -1.4 \omega$; b) $E_F = t$ (resonance with highest level)}
\end{figure}
Consider first the case where the Fermi level of the substrate is located in between
the two molecular levels (top part of Fig. \ref{noise_2_dots}).
For sharp levels (small coupling to the substrate), the noise 
has a staircase structure, as expected from photo-assisted transport, 
but contrary to the single dot case, the steps 
are not monotonous. The noise has minima close to the location of the bonding and 
antibonding orbitals. When the levels are broadened by a better coupling to the substrate, 
these steps are smoothed out, but the minima survive at first. Further increasing
the coupling to the substrates gives rises to monotonous steps which resemble 
the noise characteristic of a single dot with a sharp level.
This can be explained as follows. As long as the substrate coupling $\Gamma_S$ is smaller 
than the hopping between the two sites of the {\it molecule}, increasing  $\Gamma_S$
smoothes out the photo--assisted shot noise. But for $\Gamma_S> t$, the system 
behaves as if it was a single dot (the dot closest to the tip). This interpretation is confirmed by comparing the density of states of the double dots/single dot case (Eq. (\ref{density1})) : the limit of a very large broadening $\Gamma_S>> t$ for the double dots density of states is simply the single dot density of states with a width $t^2/(\pi \rho_S \Gamma_S^2) (<<1)$.
Therefore, 
increasing  $\Gamma_S$ brings sharper features in the noise rather than smoothing  them. Then, the noise associated with the large coupling in Fig.~\ref{noise_2_dots},
is quite similar to the staircase encountered in Fig.~\ref{grandac}.

Turning now to the situation where the Fermi level of the substrate coincides with the 
highest orbital of the {\it molecule}, one notices a noise characteristic which is again 
similar to a single dot level, except for the fact that it is inverted. For weak coupling 
to the substrate, one gets sharp steps, but as these steps are smoothed upon increasing the 
coupling, one recovers the same crossover back to the sharp steps of a single level. It does seem natural that if it is now the
lowest level of the {\it molecule} which is pinned to the substrate Fermi energy, the noise characteristic is inverted with respect
to Fig.~\ref{noise_2_dots}b (not shown). In particular, for a very strong substrate coupling, we recover curves similar 
to Fig.~\ref{moyenac}.


\section{Higher harmonics of the current}

Using the definition of the density of states in Eq.~(\ref{density1}),
 Eq.~(\ref{current_dressed}) gives for the {\it m}th harmonics of the current : 

\begin{multline}
\label{harm_gen}
I(\Omega = m \omega) = 2 e \pi \rho_T  |\Gamma^0_T|^2 \sum_n 
J_n \left( {\omega_1 \over \omega } \right) \\
\times \Bigg\{J_{n+m} \left( {\omega_1 \over \omega } \right)  (F(-\infty,\omega_0 + (n+m)\omega) + F(E_F, \omega_0 + n \omega) ) \\
-  J_{n-m}\left( {\omega_1 \over \omega } \right) F(-\infty,E_F) \Bigg\} \\
\end{multline}

The stationary current is given by taking $m=0$ in this expression :

\begin{eqnarray}
\label{current_stat_gen}
&& I(\Omega = 0) = 4 e \pi \rho_T  |\Gamma^0_T|^2 \sum_n 
J_n^2 \left( {\omega_1 \over \omega } \right) F(E_F,\omega_0 + n \omega)
\nonumber \\
\end{eqnarray}

According to Eq.~(\ref{noise_gen}), Shottky relation occurs between the zero frequencies noise and the stationary current : $S(0,0) = 2e |I(\Omega = 0)|$.

The density of states for the one/double dots sytem is even (Eq.~(\ref{density1})) so, according to Eq.~(\ref{harm_gen}), the harmonics of the current have symmetry properties : $I(\omega)$ and $I(3 \omega)$ are even and  $I(2\omega)$ is odd.

The harmonics of the current plotted in Figs \ref{1dot_harm} and \ref{2dot_harm} are normalized by $2 e \pi \rho_T |\Gamma^0_T|^2$.

\subsection{Single dot problem}

The behavior of the finite frequency higher harmonics of the current $I(m\omega)$ is examined where $\omega$ is the 
driving frequency of the AC-field. 

The stationary current is computed from Eq.~(\ref{current_stat_gen}) :

\begin{eqnarray}
&& I(\Omega = 0) = 4 e \pi \rho_T  |\Gamma^0_T|^2 \sum_n 
J_n^2 \left( {\omega_1 \over \omega } \right)\nonumber \\
&&~ \left[
\arctan\left(
{(E_D - E_F) \over \pi \rho_S  \Gamma_S^2}
\right)
- \arctan\left(
{ (E_D - \omega_0 - n \omega ) \over \pi \rho_S  \Gamma_S^2 }
\right)
  \right]\nonumber\\
\end{eqnarray}

\begin{figure}
\psfrag{omega0}{${\omega_0}/{\omega}$}
\psfrag{H0}{$I(0)$}
\psfrag{H1}{$I(\omega)$}
\psfrag{H2}{$I(2 \omega)$}
\psfrag{H3}{$I(3 \omega)$}
\centerline{\includegraphics[width=6.cm]{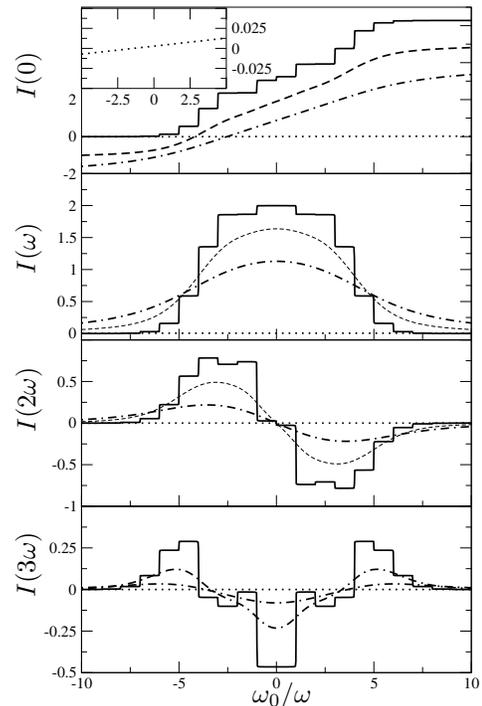}}
\caption{\label{1dot_harm} Stationary current and its three lowest harmonics as a function
 of the ratio $\omega_0 / \omega$ for the single dot system. The AC bias is fixed 
to $\omega _1 = 5.0 \omega$. The Fermi energy and the dot level  are the same than in previous 
graph ($E_F = -1.4 \omega$, $E_2 = 0.0$). Full line, $\Gamma_S = 0.001$; dashed line, $\Gamma_S = 1.0$;
dashed-dotted, $\Gamma_S = 3.0$; dotted line, $\Gamma_S = 1000.0$. Inset : stationary current as a function of $\omega_0 / \omega$ for small values of this ratio ($\Gamma_S = 1000.0$).}
\end{figure}

Fig.~\ref{1dot_harm} shows the zero-frequency component and the first three 
harmonics as a function of the ratio $\omega_0/\omega$ for different values of the substrate-{\it molecule} 
coupling which defines the effective width of the single energy level of dot, $\pi\rho_S\Gamma_S^2$. 
The zeroth harmonic represents a generalized Current-Voltage, I-V, curve. For a dot with a sharp 
level structure, the I-V curve displays a staircase structure with steps width corresponding to 
either $\Delta\omega_0=\omega$ or $\Delta\omega_0=2\omega$ associated with photo-assisted transport. 
Depending on the overlap between the Fermi distribution of the substrate and the band-width of the 
impurity the zeroth frequency harmonic can be either positive or negative leading to current inversion 
at large negative DC bias. The Ohmic regime is recovered for large $\Gamma_S$ which corresponds to 
the physical limit of a metal (See inset on Fig.~\ref{1dot_harm}a).

A general feature of the higher harmonics is the presence of maxima and minima   whose number increases 
with the order of the harmonic. These oscillations are damped as $\Gamma_S$ increases and the higher 
order responses vanish while the zeroth harmonic survives. This has important implications for the 
experiment suggested in reference \cite{kochanski}, because large $\Gamma_S$ mimics in our 
model a metallized dot, precisely the situation where Kochanski suggested there would be no higher-order 
signal. This also agrees with the results on insulating surfaces and stresses the importance of exploring 
the use of higher harmonics for more general {\it molecule}-electrode interfaces.

\vskip1cm 

\subsection{Two--site molecule}

\vskip1cm 

The double dots current harmonics are obtained from Eq. (\ref{harm_gen}) using the appropriate $F$ function (see Eqs. (\ref{density1}) and (\ref{F}))

\begin{figure}
\psfrag{omega0}{${\omega_0}/{\omega}$}
\psfrag{H0}{$I(0)$}
\psfrag{H1}{$I(\omega)$}
\psfrag{H2}{$I(2 \omega)$}
\psfrag{H3}{$I(3 \omega)$}
\centerline{\includegraphics[width=6.cm]{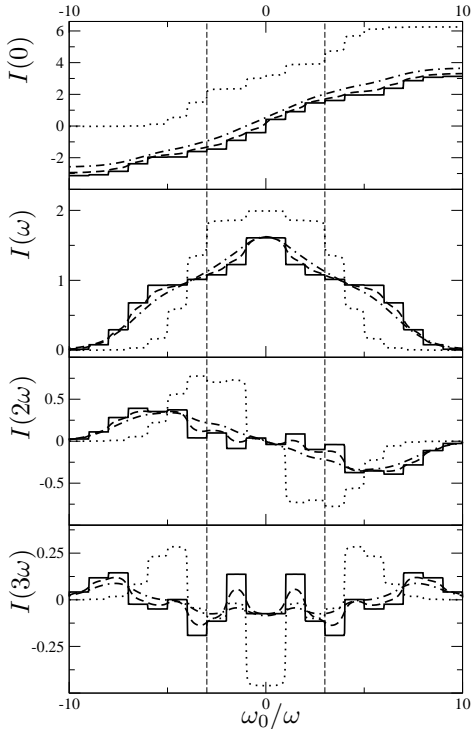}}
\caption{\label{2dot_harm} Stationary current and its three lowest harmonics as a function
 of the ratio $\omega_0 / \omega$ for the two site {\it molecule}. The alternative voltage is fixed 
to $\omega _1 = 5.0 \omega$. The Fermi energy $E_F = -1.4 \omega$ and $t = 3.0$. 
Full line, $\Gamma_S = 0.001$; dashed line, $\Gamma_S = 1.0$;
dashed-dotted, $\Gamma_S = 3.0$; dotted line, $\Gamma_S = 1000.0$. }
\end{figure}

Turning now to the discussion of the current harmonics for the 2 dot  system (Fig.~\ref{2dot_harm}), one first notices -as observed in the single dot case-
that higher order harmonics oscillate more. For comparison with the previous case, the 2 dot levels have been chosen
symmetrically around the position of the single dot level in Fig.~\ref{1dot_harm}.

 Starting with sharp levels (poor coupling
to the substrate), the zeroth harmonics has a monotonous staircase structure, ranging from negative to positive value. The first 
harmonic is positive, as in the one dot case, and has a broader width due to the spacing between the two levels. For the next
 two harmonics, the oscillatory behaviour is more prononced than in the one dot case. As in the single dot case, the harmonics have symmetry
properties : $I(\omega)$ is even,  $I(2\omega)$ is odd, $I(3 \omega)$ is even.

Upon increasing the levels widths, the amplitude of the oscillations first decreases (up to $\Gamma_S \sim t$).
For very large coupling (dotted curve on the figure), we recover exactly the one dot case : as explained before, for large $\Gamma_S$, the density of states of the double site {\it molecule} (Eq.~(\ref{density1})) corresponds to the density of states in the single dot case for a very small broadening $t^2/(\pi \rho_S \Gamma_S^2)$.
\vskip1cm 

\section{Conclusions}

In this paper, we have studied the photo-assisted shot noise for a simple model of 
metal-{\it molecule} substrate junction which could prove to be relevant in alternating current STM
experiment. The Keldysh method allowed to treat the time-dependent response of this system and in 
particular to express the noise in terms of the dressed Green's functions of the {\it molecule} plus substrate compound.
Contrary to scattering formulations of transport, the Keldysh approach allows to take into account the {\it molecule} 
occupancy correctly.
While the present deals with non-interacting electrons throughout, the general approach allows to tackle
more complicated systems where the {\it molecule} would include electron-electron interactions as they occur in carbon 
nanotubes or more complex nano-objects.

So far, the work on photo-assisted shot noise has focused mostly on mesoscopics conductors where the 
effective transmission coefficient has little energy level structure\cite{adeline}. One of the purpose of this work was
precisely to go beyond this limitation. In fact, we have seen that the presence of the {\it molecule} levels
modifies drastically the staircase which is observed in the noise derivative as a function of bias. 
As the coupling to the substrate is increased, the sharp features which we observed get more pronounced.

This corresponds to a poorly conducting system (insulator). It is relevant to follow Kochansky's intuition when 
analyzing the behaviour of the current harmonics as a function of this coupling. Indeed, the harmonics 
stay sharp in the case of an insulator. To our knowledge, this is the first time that this is addressed in a
quantitative manner, albeit with the help of a simple model. 
The fact that the $I(3\omega)$ signal has been
measured for insulating surfaces makes it reasonable to assume that the experiment with
electrode-{\it molecule} junctions will soon be done.  
It is on this situation that the predictions of this model may be
more relevant. 

Future directions of research could include a time-dependent, self-consistent treatment of Coulomb interactions
in the {\it molecule}. Electronic correlations has recently been addressed in References~\cite{thielmann,oguri,asano}.
Another possible extension would be to describe the {\it molecule} using {\it ab initio} method, as used in~\cite{baer,taylor,brandbyge} 
where the effect of an AC bias in molecular transport is also addressed.

\acknowledgements
The authors wish to thank Philippe Dumas for useful discussions. V. M. would like to acknowledge CNRS for funding his stay in Marseille.

\appendix

\section{Unperturbed Green's functions}

\subsection{For two electrodes}

\label{appelectrode}

The non--equilibrium Green's functions for a site $M$ ($M = T,S$) into a  normal metal electrode are:

\begin{eqnarray}
g^{+-}_{MM}(\Omega) &=& 2 i \pi \rho_M f(\Omega) \nonumber \\
g^{-+}_{MM}(\Omega) &=& - 2 i \pi \rho_M [1-f(\Omega)] \nonumber \\
g^a_{MM} (\Omega) &=& i \pi \rho_M = g^{r *}_{MM} (\Omega)
\end{eqnarray}

where $f(\Omega)$ is the Fermi distribution and $\rho_M$ is the local density of states on site $M$ near the Fermi level.

\subsection{For one dot}

\label{appA}

Using the definitions of Keldysh Green's functions in terms of creation/annihilation operators of electrons in the dot, we find the temporal (homogeneous) non--equilibrium Green's functions:

\begin{eqnarray}
g^{+-}_{DD}(\tau) &=& i n^0_D e^{-i E_D \tau} \nonumber \\
g^{-+}_{DD}(\tau) &=& -i (1 -n^0_D) e^{-i E_D \tau} \nonumber \\
g^{a}_{DD}(\tau) &=& i \Theta(- \tau) e^{-i E_D \tau} \nonumber \\
g^{r}_{DD}(\tau) &=& -i \Theta(\tau) e^{-i E_D \tau} \nonumber \\
 \end{eqnarray}

where $n^0_D$ is the number electron in the dot at time $0$, $E_D$ is the energy and $\Theta(t)$ is the Heaviside function .

Using a Fourier transform, the expressions of the Green's functions in the frequency space look:

\begin{eqnarray}
g^{+-}_{DD}(\Omega) &=& 2 i n^0_D \frac{\eta}{(\Omega -E_D)^2 + \eta^2} \nonumber \\
g^{-+}_{DD}(\Omega) &=& - 2 i (1 -n^0_D) \frac{\eta}{(\Omega -E_D)^2 + \eta^2} \nonumber \\
g^{a}_{DD}(\Omega) &=& \frac{1}{\Omega -E_D - i \eta} \nonumber \\
g^{r}_{DD}(\Omega) &=& \frac{1}{\Omega -E_D + i \eta} \nonumber \\
 \end{eqnarray}

where $\eta$ is an infinitesimal positive constant.

\subsection{For a linear system of $D$ dots ($D>1$)}

\label{appB}

A linear chain of $D$ dots is considered. The sites are labelled by $j = 1..D$. The real hopping between two nearest neighbor sites is $t$. In such a system, eigen-energies and eigenfunctions of the site $j$ are given by:

\begin{eqnarray}
\left \lbrace
\begin{array}{cc}
E_j & = - 2 t \cos\left( {\pi j \over D+1} \right) \nonumber \\
\Psi_j(l) & = K_j \sin \left( {\pi j l \over D+1} \right)
\end{array}
\right.
\end{eqnarray}

where $j,l = 1..D$. Eigenfunctions are normalized with $K_j$ given by: 

\begin{equation}
|K_j|^2 = \left[\sum_{l=1}^{D} \sin^2\left( {\pi j l \over D+1}\right)
\right]^{-1}
\end{equation}

The non--equilibrium Green's functions can be derived using the previous expressions. In position and energy space, that yields:

\begin{eqnarray}
\label{apm}
g^a(l_1,l_2,E) &=& \sum_{n=1}^{D} |K_n|^2 \frac{
\sin \left( {\pi n l_1 \over D+1} \right)\sin \left( {\pi n l_2 \over D+1} \right)}{E + 2 t \cos \left( {\pi n \over D+1} \right) -i \eta} \nonumber \\
g^{+-}(l_1,l_2,E) &=& 2 i \pi \sum_{n=1}^{D} |K_n|^2 
\sin \left( {\pi n l_1 \over D+1} \right)\sin \left( {\pi n l_2 \over D+1} \right) \nonumber \\
&&~~~~~~~~~~\times \chi(E_n) \delta(E - E_n)
\end{eqnarray}

where $ \chi(E)$ is the occupation number of energy level $E$ ($0$ or $1$) and $\eta$ is an infinitesimal positive constant. The retarded Green's function is the complex conjugate of the advanced one. The $-+$ Green's function is obtained by replacing $\chi(E_n)$ by $-(1-\chi(E_n))$ in the $+-$ expression.

In this proposal, only the case $D=2$ is studied. The expressions of the normalization constants are $|K_1|^2 = |K_2|^2 = 2/3$. The eigenfunctions and eigenvalues for each dot are given by:

\begin{eqnarray}
\left \lbrace
\begin{array}{cc}
E_1 = -t  &\;\;\;\;\;\;  E_2 = t \\
\Psi_1(l) = \sqrt{2 \over 3} \sin \left( {\pi l \over 3} \right) &\;\;\;\;\;\; \Psi_2(l) = \sqrt{2 \over 3} \sin \left( {2 \pi l \over 3} \right)
\end{array}
\right.
\end{eqnarray}

Introducing these expressions in (\ref{apm}) yields:

\begin{eqnarray}
g^a_{l_1,l_2}(E) &=& {2 \over 3} \left[{ \sin\left({\pi l_1 \over 3} \right)\sin\left({\pi l_2 \over 3} \right) \over E + t -i \eta} + { \sin\left({2 \pi l_1 \over 3} \right)\sin\left({2 \pi l_2 \over 3} \right) \over E - t -i \eta}\right] \nonumber \\
g^{+-}_{l_1,l_2}(E) &=& {4 i \pi \over 3} \Big[\sin\left({\pi l_1 \over 3} \right)\sin\left({\pi l_2 \over 3} \right) \chi(-t) \delta(E+t)\nonumber \\
&&~~~~+  \sin\left({2 \pi l_1 \over 3} \right)\sin\left({2 \pi l_2 \over 3} \right) \chi(t) \delta(E-t)  \Big] 
\end{eqnarray}

For the two dots system, only the Green's functions for $l_1,l_2 = 1,2$ are needed:

\begin{eqnarray}
g^a_{11}(E) &=& g^a_{22}(E) = {1 \over 2} \left[ {1 \over E+t -i \eta} + {1 \over E- t -i\eta}\right] \nonumber \\
g^a_{12}(E) &=& g^a_{21}(E) = {1 \over 2} \left[ {1 \over E+t -i \eta} - {1 \over E- t -i\eta}\right] \nonumber \\
g^{+-}_{11}(E) &=& g^{+-}_{22}(E) = i \pi \left[ \chi(-t) \delta(E+t) + \chi(t) \delta(E-t)
\right] \nonumber \\
g^{+-}_{12}(E) &=& g^{+-}_{21}(E) = i \pi \left[ \chi(-t) \delta(E+t) - \chi(t) \delta(E-t)
\right] \nonumber \\
\end{eqnarray}

\section{Dressed Green's functions}

In this section, dressed Green's functions that enter in the expression of the current and noise are derived. In the general case of a linear chain of $N$ dots, it is possible to compute the dressed Green's functions using a linear set of Dyson equations. As explained in the previous section, the usual scheme is to isolate the sub--system composed by the substrate and the chain and then to express the dressed Green's functions of the point contacts in term of the unperturbed Green's functions of the isolated chain and of the substrate. Standard unperturbed Green's functions of a metallic electrode are used for the substrate (Appendix  \ref{appelectrode}) including the Fermi energy $E_F$ and the density of states $\rho_S$  of the substrate. 

\subsection{For one dot}

\label{appdressed}

Here, only the system composed by the substrate ($S$) and one dot is considered. The hopping between the substrate and the {\it molecule} will be denoted by $\Gamma_S$ and will be chosen to be real. The dressed Green's functions have homogeneity in time due to the time independence of the hopping Hamiltonian. It also permits to use directly the Dyson equation in frequency. 

For the advanced and retarded Green's function, using two Dyson equations rises:

\begin{eqnarray}
\label{dress1}
\tilde{G}^{a,r}_{DD}(\Omega) &=&  g^{a,r}_{DD}(\Omega) +  g^{a,r}_{DD}(\Omega) \Sigma^{a,r}_{DS} \tilde{G}^{a,r}_{SD}(\Omega) \nonumber \\
&=& \frac{g^{a,r}_{DD}(\Omega)}{ 1 \mp i \pi \rho_S \Gamma_S^2 g^{a,r}_{DD}(\Omega)}
\end{eqnarray} 

where $\Sigma^{a,r}_{DS} = \Gamma_S$ the self energy associated to the hopping between the substrate and the dot and the upper/downer sign stands for the advanced/retarded Green's functions. Little $g$ denote the unperturbed non--equilibrium Green's functions for an isolated dot. Their expressions are given in Appendix \ref{appA}.  
Introducing these expressions in the dressed Green's function of the point contact gives:

\begin{eqnarray}
\tilde{G}^{a,r}_{DD}(\Omega) &=&  \frac{1}{ \Omega -E_D \mp i(\eta +\pi \rho_S \Gamma_S^2)}
\end{eqnarray} 

where $\eta$ is an infinitesimal positive constant. This expression show the effect of the substrate. Due to the hopping $\Gamma_S$ between the dot and the electrode, a broadening of the energy level of the dot appears.  

The limit when $\eta$ tends to $0$ is straightforward because these Green's functions have no real poles:

\begin{eqnarray}
\tilde{G}^{a,r}_{DD}(\Omega) &=&  \frac{1}{ \Omega -E_D \mp i   \pi \rho_S  \Gamma_S^2}
\end{eqnarray}

The same technique allows to obtain the spectral dressed Green's functions non--perturbatively:

\begin{eqnarray}
\label{dress2}
\tilde{G}^{+-}_{DD}(\Omega) &=& \frac{
g^{+-}_{DD} (\Omega) + 2 i \pi \rho_S  \Gamma_S^2 |g^a_{DD}(\Omega)|^2 f(\Omega-E_F)
}{|1 - i \pi \rho_S  \Gamma_S^2 g^a_{DD}(\Omega)
|^2} \nonumber \\
\tilde{G}^{-+}_{DD}(\Omega) &=&  \frac{
g^{-+}_{DD} (\Omega) + 2  i  \pi \rho_S  \Gamma_S^2 |g^a_{DD}(\Omega)|^2 (1 - f(\Omega-E_F))
}{|1 - i  \pi \rho_S  \Gamma_S^2 g^a_{DD}(\Omega)
|^2} \nonumber \\
\end{eqnarray}

Taking the limit when $\eta \rightarrow 0$ gives:

\begin{eqnarray}
\tilde{G}^{+-}_{DD}(\Omega) &=& 
2 i \pi \rho_S  \Gamma_S^2
\frac{ f(\Omega-E_F)
}{
(\Omega - E_D)^2 +( \pi \rho_S  \Gamma_S^2
)^2} \nonumber \\
\tilde{G}^{-+}_{DD}(\Omega) &=& - 2 i  \pi \rho_S  \Gamma_S^2
\frac{ (1 - f(\Omega-E_F))
}{
(\Omega - E_D)^2 +( \pi \rho_S  \Gamma_S^2
)^2} \nonumber \\
\end{eqnarray}

\subsection{For a linear system of $D$ dots ($D>1$)}

\label{appdressed1}

In order to compute the dressed Green's functions in a non--perturbative way between the substrate and the linear chain of dots, the system is studied in the absence of the tip. The $D$ dots are labelled by $1..D$, where the site $1$ is the point contact between the substrate and the chain and the site $D$ is the point contact between the chain and the tip. Green's functions for a linear chain of dots are derived in the appendix \ref{appB} and will be denoted by a little $g$. Here, the relevant non--equilibrium Green's functions are those for extreme dots.

In this system, the hopping Hamiltonian between the substrate and the chain is time--independent (and so, the Green's functions are homogeneous in time), it permits to use directly the frequency expression of Dyson equation. In the noise and current formulae, only the dressed Green's function for the site $D$ is needed. To express in a non--pertubative scheme these Green's functions, one needs to write a linear set of Dyson equations. Here is the expression of the advanced and retarded Keldysh Green's functions for the site $D$:

\begin{equation}
\label{GFar}
\tilde{G}^{r,a}_{DD}(\Omega) = g^{r,a}_{DD}(\Omega) +  g^{r,a}_{D1}(\Omega) \Sigma^{a,r}_{1S} \tilde{G}^{r,a}_{SD}(\Omega)
\end{equation}

where $ \Sigma^{a,r}_{1S} =\Gamma_S$ is the self--energy for a single hopping and $\tilde{G}^{r,a}_{SD}(\Omega)$ is a dressed Green's function that has to be expand by a new Dyson equation. $g^{r,a}_{D1}(\Omega)$ contains all the informations of the isolated linear chain.
We find a non--perturbative expression for this Green's function to all orders of $\Gamma_S$:

\begin{equation}
\tilde{G}^{r,a}_{SD}(\Omega) = \frac{\mp i \pi \rho_S \Gamma_S g^{r,a}_{1D}(\Omega)}{1 \pm  i \pi \rho_S \Gamma_S^2 g^{r,a}_{11}(\Omega)}
\end{equation}

So a formal expression for the non--perturbative dressed advanced and retarded Green's functions are obtained  in terms of the unperturbed Green's functions of the linear chain of dots:

\begin{equation}
\tilde{G}^{r,a}_{DD}(\Omega) = g^{r,a}_{DD}(\Omega) \mp i \pi \rho_S \Gamma_S^2 \frac{g^{r,a}_{1D}(\Omega) g^{r,a}_{D1}(\Omega)}{1 \pm  i \pi \rho_S \Gamma_S^2  g^{r,a}_{11}(\Omega)}
\end{equation}

For the spectral Green's functions, the same technique is applied and permits also to have an expression for the dressed Green's functions:

\begin{eqnarray}
\label{GFpmmp}
&& \tilde{G}^{+-}_{DD}(\Omega) = g^{+-}_{DD}(\Omega) \nonumber \\
&& + i \pi \rho_S \Gamma_S^2 \left(
\frac{g^{a}_{1D}(\Omega) g^{+-}_{D1}(\Omega)}{1 -  i \pi \rho_S \Gamma_S^2 g^{a}_{11}(\Omega)} - \frac{g^{r}_{D1}(\Omega) g^{+-}_{1D}(\Omega)}{1 +  i \pi \rho_S \Gamma_S^2 g^{r}_{11}(\Omega)}
\right)\nonumber \\
&&+ \pi^2 \rho_S^2 \Gamma_S^4 \frac{g^a_{1D} (\Omega)   g^{r}_{D1} (\Omega)    g^{+-}_{11}  (\Omega)
}{
(1 +  i \pi \rho_S \Gamma_S^2 g^{r}_{11}(\Omega))(1 -  i \pi \rho_S \Gamma_S^2 g^{a}_{11}(\Omega))
}\nonumber\\
&& + 2 i \pi \rho_S f(\Omega - E_F) \Gamma_S^2 g^r_{D1}(\Omega)  g^a_{1D}(\Omega) \nonumber \\
&&~~~ \times \left[1 - i\pi \rho_S \Gamma_S^2 \frac{g^r_{11} (\Omega)}{1+  i\pi \rho_S \Gamma_S^2 g^r_{11} (\Omega)} \right] \nonumber \\
&&~~~ \times \left[1 + i\pi \rho_S \Gamma_S^2 \frac{g^a_{11} (\Omega)}{1-  i\pi \rho_S \Gamma_S^2 g^a_{11} (\Omega)} \right]  \nonumber \\
&& \tilde{G}^{-+}_{DD}(\Omega) = g^{-+}_{DD}(\Omega) \nonumber \\
&& + i \pi \rho_S \Gamma_S^2 \left(
\frac{g^{a}_{1D}(\Omega) g^{-+}_{D1}(\Omega)}{1 -  i \pi \rho_S \Gamma_S^2 g^{a}_{11}(\Omega)} - \frac{g^{r}_{D1}(\Omega) g^{-+}_{1D}(\Omega)}{1 +  i \pi \rho_S \Gamma_S^2 g^{r}_{11}(\Omega)}
\right)\nonumber \\
&&+ \pi^2 \rho_S^2 \Gamma_S^4 \frac{g^a_{1D} (\Omega)   g^{r}_{D1} (\Omega)    g^{-+}_{11}  (\Omega)
}{
(1 +  i \pi \rho_S \Gamma_S^2 g^{r}_{11}(\Omega))(1 -  i \pi \rho_S \Gamma_S^2 g^{a}_{11}(\Omega))
}\nonumber\\
&& - 2 i \pi \rho_S (1 - f(\Omega -E_F))\Gamma_S^2 g^r_{D1}(\Omega)  g^a_{1D}(\Omega)\nonumber \\
&&~~~ \times  \left[1 - i\pi \rho_S \Gamma_S^2 \frac{g^r_{11} (\Omega)}{1+  i\pi \rho_S \Gamma_S^2 g^r_{11} (\Omega)} \right] \nonumber \\
&&~~~\times \left[1 + i\pi \rho_S \Gamma_S^2 \frac{g^a_{11} (\Omega)}{1-  i\pi \rho_S \Gamma_S^2 g^a_{11} (\Omega)} \right]  \nonumber \\
\end{eqnarray}

In this paper only the case of a double dot {\it molecule} is considered ($D = 2$) . The unperturbed Green's functions for a double dot system are derived in appendix \ref{appB}. Introducing these expressions in Eqs.(\ref{GFar}) and (\ref{GFpmmp}) gives:

\begin{eqnarray}
\tilde{G}^{a}_{DD}(\Omega) &=& -  \frac{ \Omega - i\pi \rho_S \Gamma_S^2} {t^2 - \Omega (\Omega -  i \pi \rho_S \Gamma_S^2)} \nonumber \\
\tilde{G}^{+-}_{DD}(\Omega) &=& 2 i \pi \rho_S \Gamma_S^2\frac{t^2 f(\Omega - E_F)}{|t^2 - \Omega (\Omega +  i \pi \rho_S \Gamma_S^2)|^2} 
\end{eqnarray}

where $t$ is the hopping between the dots. The dressed retarded Green's function is the complex conjugate of the advanced one and the $-+$ Green's function is obtained by replacing $f(\Omega -E_F)$ in the $+-$ expression by $-(1 - f(\Omega - E_F))$.

\end{multicols}
\end{document}